\documentclass[twocolumn,showpacs,preprintnumbers,amsmath,amssymb]{revtex4}
\usepackage{amsmath}
\usepackage{amssymb}
\usepackage{graphicx}
\usepackage{bm}
\usepackage{epstopdf}

\begin{document}

\title{Field theoretic description of partially reflective surfaces}

\author{F. A. Barone}
 \email{fbarone@unifei.edu.br} 
\affiliation{ICE - Universidade Federal de Itajub\'a, Avenida BPS 1303, Caixa Postal 50, CEP. 37500-903, Itajub\'a, MG, Brazil.}
\author{F. E. Barone}
 \email{febarone@cbpf.br} 
\affiliation{Centro Brasileiro de Pesquisas F\'\i sicas, Rua Dr.\ Xavier Sigaud 150, Urca, CEP.  22290-180, Rio de Janeiro, RJ, Brazil.}

\date{\today}

\begin{abstract}
The issue of electric charges in interaction with partially reflective surfaces is addressed by means of field theoretic methods. It is proposed an enlarged Maxwell lagrangian, describing the electromagnetic field in the presence of a semitransparent surface, and its corresponding photon propagator is computed exactly. The amended Green function reduces to the one for a perfect conductor in the appropriate limit, and leads to the interaction between charges and surfaces with varying degrees of transparency, featured by a phenomenological parameter. The interaction found via image method is recovered, in the limiting case of perfect mirrors, as a testimony to the validity of the model.
\end{abstract}

\maketitle

\section{Introduction}

The empty scenario underlying the processes in classical physics, has acquired a rich structure in quantum field theory. The vacuum is now the arena where the interaction potential takes place by means of the exchange of virtual particles, which are characterized by their propagators; therefore, its functional form determines the interaction energy of the system.

Modifications in the photon propagator due to boundary conditions have been a long-standing issue in regards to the Casimir effect (see for instance \cite{Miltonlivro,Bordag}, and references cited therein). The first field theoretic treatment of this effect for infinitesimally thin perfect conductors was devised in \cite{BRW}. Recently \textit{Parashar}, \textit{Milton}, \textit{Shajesh} and \textit{Schaden} carried out a detailed analysis clarifying the controversies involving different models of boundary condition and studied the Casimi-Polder interaction between an atom and a $\delta$-function plate, among other things \cite{Parashar}. This important issue, of atoms and charges in interaction with material boundaries, has generated a renewed interest for both practical and theoretical reasons \cite{ABFT,Milton2,Bordag2,Eberlein1}. In this context \textit{Eberlein} and \textit{Zietal} devised a remarkable method \cite{Eberlein2}, closely related to the electrostatic image method \cite{Farina}, to deal with a wide range of physical setups.

Besides, $\delta$-like potentials coupled to quantum fields have been widely used to describe partially reflective surfaces in order to investigate the Casimir effect \cite{Miltonlivro,Bordag,Milton,esferadelta,delta1d,Ricardo,softMit}.  This same kind of potential was also used to show that the scalar boson propagator undergoes a modification in such a way that the interaction between a point charge and the surface of an infinitesimally thin mirror becomes equal to the one obtained by employing the image method \cite{CBB}. In spite of their success, these methods cannot be used to describe the behaviour of the electromagnetic field in the presence of a reflective surface due to its gauge invariance.

To overcome this challenge, in this paper we propose an enlarged Maxwell lagrangian which can describe the electromagnetic field in the presence of a $\delta$-like semitransparent mirror.  This is achieved by introducing a new term that modifies the propagator according to this anisotropic setup. A similar approach has been developed by \textit{Fosco}, \textit{Lombardo} and \textit{Mazzitelli} in \cite{FLM}, where the authors couple the gauge field to the mirror setting a new term in the action.

Specifically, in this work we deal with a vector field $A_{\mu}$ in $(1+3)$ dimensions and spacetime metric $\eta_{\mu\nu}=\textrm{diag }(+,-,-,-)$. In section (\ref{propagator}) we define an amended Maxwell lagrangian, adding a new term suitable to describe a $\delta$-like uniaxial dielectric surface, and we find the modification undergone by the free photon propagator due to the presence of this term. The interaction between the surface and a point charge is investigated in section (\ref{image}). The general result obtained turns out to be the exact analytical expression for the interaction energy between pointlike charges and the semitransparent mirror, explicitly exhibiting its dependence on the surface degree of transparency. Section (\ref{final}) is devoted to final remarks.

\section{The modified photon propagator}
\label{propagator}

To model the electromagnetic field in the presence of a $\delta$-like mirror we start off by imposing, in an \textit{ad hoc} manner, an additional term in the Maxwell lagrangian with the usual covariant gauge fixing. From now on, and with no loss of generality, we will consider a semitransparent plate perpendicular to the $x^{3}=z$ axis and located at position ${\bf a}=(0,0,a)$; therefore, the normal vector to the surface is $S^{\gamma}=\eta^{\gamma}_{\ 3}$, and the model reads,
\begin{equation}
\label{EQ1}
\mathcal L = -\frac{1}{4}(F)^{2}-\frac{1}{2\xi}(\partial A)^2-\frac{1}{m} \Biggl(\frac{1}{2}S^{\mu}\epsilon_{\mu\nu\alpha\beta}F^{\alpha\beta}\Biggr)^{2}\delta(x^{3}-a) \ .
\end{equation}
Note that, as it stands in the lagrangian, it is the dual field-strength tensor, $\widetilde{F}_{\mu\nu}=\frac{1}{2}\epsilon_{\mu\nu\alpha\beta}F^{\alpha\beta}$, that is contracted with the  normal vector  to the surface. 

The constant $m\geq0$, has dimension of mass in natural units and is introduced as a measure of the mirror degree of transparency. That it is a function of the electromagnetic properties of the material can be clearly seen from its relation to the electric permitivity,  $\epsilon^{ij}$, and inverse magnetic permeability, $(\mu^{-1})^{ij}$, of the model (\ref{EQ1}),
\begin{eqnarray}
\label{EQ1.1}
\epsilon^{ij} &=& \delta^{ij}+\frac{2}{m} \delta(x^{3}-a)(\delta^{i1} \delta^{j1}+ \delta^{i2} \delta^{j2}) \ , \cr\cr
(\mu^{-1})^{ij} &=& \delta^{ij}+\frac{2}{m} \delta(x^{3}-a)(\delta^{i3} \delta^{j3}) \ .  
\end{eqnarray}
These are the same kind of electromagnetic properties used in \cite{Parashar} to analyze semitransparent $\delta$-function plates.
Accordingly, the principal susceptibilities $\chi^{ii}$ that can be read off from the first equation in (\ref{EQ1.1}) show that the lagrangian (\ref{EQ1}) describes an uniaxial dielectric surface. 
 
It is worth mentioning that because the Levi-Civita tensor is totally antisymmetric, the derivatives in the last term in (\ref{EQ1}) are taken only in the parallel space to the surface, that is; 
\begin{equation}
\label{EQ1.3}
\left(\frac{1}{2}\  \eta^{\mu}_{\ 3}\epsilon_{\mu\nu\alpha\beta}F^{\alpha\beta} \right)^{2} = \epsilon_{3\alpha\beta\nu}\ \epsilon_{3\rho\tau}^{\ \ \ \ \nu} (\partial_\parallel^{\alpha}A^{\beta})(\partial_\parallel^{\rho}A^{\tau}) \ , \nonumber
\end{equation}
where we defined the differential operator $\partial_\parallel^{\alpha}=(\partial^{0},\partial^{1},\partial^{2},0)$.

The model exhibits a $\delta$-type discontinuity on the mirror, and the influence of this discontinuity on the fields can readily be understood from the equation of motion,
\begin{equation}
\label{EQ.MOV}
\partial^{\mu} F_{\mu\nu} + \frac{2}{m} \delta(x^{3}-a) \mathcal V_{\nu} = J_{\nu}\ ,
\end{equation}
where,
\begin{equation}
\label{EQ.V}
\mathcal V_{\nu}= \partial^{\alpha} F_{\alpha\nu} + \partial_{3} F_{3 \nu} + \eta_{3 \nu} \partial^{\alpha} F_{\alpha 3} \ .
\end{equation}
It is this vector that dictates the behavior of the fields exactly on the boundary $x^{3}=a$. This subject can be clarified by writing Eq.(\ref{EQ.MOV}) for $\nu=0,1,2,3$, which leads to
\begin{eqnarray}
\label{EQ.V1}
\nabla\cdot\Bigl[\textbf{E}+\frac{2}{m}\delta(x^{3}-a)\textbf{E}_{\|}\Bigr]&=& J_{0} \ ,\cr\cr
\nabla\times\Bigl[\textbf{B}+\frac{2}{m}\delta(x^{3}-a)\textbf{B}_{\perp}\Bigr]&=& \textbf{J} \cr\cr 
+\frac{\partial}{\partial t}\Bigl[&\textbf{E}&+\frac{2}{m}\delta(x^{3}-a)\textbf{E}_{\|}\Bigr]
\end{eqnarray}
where we defined the vectors perpendicular and parallel to the plate, namely, $\textbf{E}_{\|}=(E^{1},E^{2},0)$, $\textbf{B}_{\perp}=(0,0,B^{3})$.

Using Eqs.(\ref{EQ1.1}) we have
\begin{eqnarray}
\label{def:D,H}
D^{i}=\sum_{j}\epsilon^{ij}E^{j}&\Rightarrow&{\bf D}={\bf E}+\frac{2}{m}\delta(x^{3}-a){\bf E}_{\|}\ ,\cr\cr
H^{i}=\sum_{j}(\mu^{-1})^{ij}B^{j}&\Rightarrow&{\bf H}={\bf B}+\frac{2}{m}\delta(x^{3}-a){\bf B}_{\perp}\ ,
\end{eqnarray}
so that Eqs.(\ref{EQ.V1}) can be rewritten in the form
\begin{eqnarray}
\label{def:P,M}
\nabla\cdot{\bf D}=J^{0}\ \ ,\ \ \nabla\times{\bf H}={\bf J}+\frac{\partial{\bf D}}{\partial t}
\end{eqnarray}

Once the polarization and magnetization vectors are defined by ${\bf D}={\bf E}+{\bf P}$ and ${\bf H}={\bf B}-{\bf M}$, respectively, from (\ref{def:D,H}) we see that 
\begin{eqnarray}
{\bf P}=\frac{2}{m}\delta(x^{3}-a){\bf E}_{\|}\ \ ,\ \ 
{\bf M}=-\frac{2}{m}\delta(x^{3}-a){\bf B}_{\perp}\ ,
\end{eqnarray}
showing that the $\delta$-type discontinuities are entirely contained in the polarization and magnetization vectors, defined only on the plate.

Despite the $\delta$-type discontinuity it turns out that the propagator is well defined all over the space, as we will see. This is closely related to the absence of any kind of \textit{ad hoc} boundary condition, in the sense that we do not require any constraint on the gauge fields to describe the material surface. The new term in (\ref{EQ1}) is enough to set the necessary conditions at $x^{3}=a$. The same approach has been used in \cite{FLM} where the authors start from an action that includes an interaction term coupling the gauge field to the mirror, without resorting to \textit{ad hoc} boundary conditions.

By setting the Feynman gauge $(\xi=1)$ in (\ref{EQ1}) and bringing it out to the usual quadratic form we can find the operator we are interested in,
\begin{eqnarray}
\label{EQ2}
\mathcal L &=&\frac{1}{2}A_{\mu}\mathcal O^{\mu\nu} A_{\nu} \cr\cr
&=& \frac{1}{2}A_{\mu}\left[\eta^{\mu\nu}\square +\frac{2}{m}\ \delta(x^{3}-a)(\eta^{\mu\nu}\square_\parallel-\partial_\parallel^{\mu}\partial_\parallel^{\nu})\right]A_{\nu}  \ , 
\end{eqnarray}
where $\square_\parallel=\partial_\parallel^{\alpha}{\partial_\parallel}_{\alpha}$.

For the sake of simplicity in the calculations that follow, let us split the above differential operator into two parts: 
\begin{equation}
\label{EQ3}
\mathcal O^{\mu\nu} = \mathcal O^{(0)\mu\nu} + \Delta \mathcal O^{\mu\nu}  \ , 
\end{equation}
where,
\begin{eqnarray}
\label{EQ4}
\mathcal O^{(0)\mu\nu} &=& \eta^{\mu\nu}\square \ , \cr\cr
\Delta \mathcal O^{\mu\nu} &=& \frac{2}{m} \ \delta(x^{3}-a)(\eta^{\mu\nu}\square_\parallel-\partial_\parallel^{\mu}\partial_\parallel^{\nu})  \ . 
\end{eqnarray}
We will also write $G^{(0)\mu\nu}$ for the free photon propagator that, in coordinate space, is defined by the relation $\mathcal O^{(0)\mu\nu}(x) G^{(0)}_{\nu\lambda}(x,y) = \eta^{\mu}_{\ \lambda} \delta^{(4)}(x-y)$.

As in the scalar field case \cite{CBB}, the modified photon propagator $G^{\mu\nu}(x,y)$ due to the presence of the mirror, or the operator that inverts (\ref{EQ3}), can be given recursively in integral form as, 
\begin{eqnarray}
\label{EQ5}
G_{\mu\nu}(x,y) &=& G^{(0)}_{\mu\nu}(x,y)\cr\cr
&-& \int d^{4}z \ G_{\mu\gamma}(x,y) \Delta \mathcal O^{\gamma\sigma}(z) G^{(0)}_{\sigma\nu}(z,y) \ . 
\end{eqnarray}

For convenience, let us Fourier transform the Green functions in the coordinates parallel to the mirror in order to get the reduced Green functions from,
\begin{eqnarray}
\label{EQ6}
G_{\mu\nu}(x,y) = \int \frac{d^{3}p_\parallel}{(2\pi)^{3}} \ \mathcal G_{\mu\nu}(x^{3},y^{3};p_\parallel) \  e^{-ip_\parallel(x_\parallel-y_\parallel)} \ . 
\end{eqnarray}
With the above equation, the free reduced propagator is easily found as,
\begin{eqnarray}
\label{EQ7}
\mathcal G^{(0)}_{\mu\nu}(x^{3},y^{3};p_\parallel) &=& -\eta_{\mu\nu} \int \frac{dp^{3}}{2\pi} \frac{e^{ip^{3}(x^{3}-y^{3})}}{p_\parallel^{2}-(p^{3})^{2}} \cr\cr
&=& \eta_{\mu\nu} \frac{e^{-\sigma |x^{3}-y^{3}|}}{2\sigma} \ , 
\end{eqnarray}
where we defined $\sigma = \sqrt{-p^{2}_\parallel}$.

Substituting the last definition in (\ref{EQ4}) into (\ref{EQ5}), transforming the result according to (\ref{EQ6}) and using (\ref{EQ7}), after some straightforward integrations we find the modified photon propagator we are searching for,
\begin{eqnarray}
\label{EQ8}
\mathcal G_{\mu\nu}(x^{3},y^{3};p_\parallel) &=& \eta_{\mu\nu} \frac{e^{-\sigma |x^{3}-y^{3}|}}{2\sigma} \cr\cr
+ \frac{2}{m} \mathcal G_{\mu\gamma}(x^{3},a;p_\parallel) \!\! & p_\parallel^{2} & \!\!  
\big({\eta_\parallel}^{\gamma}_{\ \nu} - \frac{p_\parallel^{\gamma}{p_\parallel}_{\nu}}{p_\parallel^{2}} \big)\frac{e^{-\sigma |a^{3}-y^{3}|}}{2\sigma} \ ,
\end{eqnarray}
where we defined ${\eta_\parallel}^{\mu\nu}=\eta^{\mu\nu}+\eta^{\mu 3}\eta^{\nu 3}$ and $p_\parallel^{\gamma}=(p^{0},p^{1},p^{2},0)$.

Even though the above equation still defines the reduced propagator recursively, it is possible to find out its functional form employing an approach that is a little tricky. To accomplish this task, first note that we can exploit the fact that it explicitly exhibits its dependence on the mirror position, so that we can write the  propagator from an arbitrary point to the mirror setting $y^{3}=a$, and isolate the terms containing it, which allows us to write
\begin{eqnarray}
\label{EQ9}
\mathcal G_{\mu\gamma}(x^{3},a;p_\parallel) \Biggl[ \eta^{\gamma}_{\ \nu} &-& \frac{1}{m\sigma} p_\parallel^{2} \Biggl({\eta_\parallel}^{\gamma}_{\nu}- \frac{p_\parallel^{\gamma} {p_\parallel}_{\nu}}{p_\parallel^{2}} \Biggr) \Biggr] \cr\cr
& & = \eta_{\mu\nu} \frac{e^{-\sigma |x^{3}-a|}}{2\sigma} \ . 
\end{eqnarray}
Multiplying both sides by the operator that inverts the term in brackets yields, 
\begin{eqnarray}
\label{EQ10}
&\ &\!\!\! \mathcal G_{\mu\nu}(x^{3},a;p_\parallel) = \cr\cr
&\ &\!\!\! \frac{e^{-\sigma |x^{3}-a|}}{2\sigma} \Biggl[\eta_{\mu\nu} -\frac{1}{m}  \frac{\sigma}{(1+\frac{\sigma}{m})} \Biggl({\eta_\parallel}_{\mu\nu}- \frac{{p_\parallel}_{\mu} {p_\parallel}_{\nu}}{p_\parallel^{2}} \Biggr) \Biggr]\ .\cr
&\ &\  
\end{eqnarray}

Substituting Eq. (\ref{EQ10}) in (\ref{EQ8}) and using Eq.(\ref{EQ6}), the photon propagator modified due to the presence of the mirror assumes the form,
\begin{eqnarray}
\label{EQ11}
G_{\mu\nu}(x,y) &=& \int \frac{d^{3}p_\parallel}{(2\pi)^{3}} \ \Biggl[\eta_{\mu\nu} \frac{e^{-\sigma |x^{3}-y^{3}|}}{2\sigma} \cr\cr
&-& \frac{1}{2} \frac{e^{-\sigma (|x^{3}-a|+|y^{3}-a|)}}{m+\sigma} \Biggl({\eta_\parallel}_{\mu\nu}- \frac{{p_\parallel}_{\mu} {p_\parallel}_{\nu}}{p_\parallel^{2}} \Biggr)\Biggr] \cr\cr
& & \times \  e^{-ip_\parallel(x_\parallel-y_\parallel)} \ . 
\end{eqnarray}

It is noteworthy that this propagator is continuous and well defined all over the space, as can readily be seen. As an important check we point out that, by placing one of the plates at infinity in the propagator that describes the Casimir effect for perfect conductors \cite{BRW}, we get the propagator (\ref{EQ11}) with $m=0$ (after the transformation $\Gamma = i \sigma$ just to make the notation equivalent). This crucial validation shows that both propagators are the same in the appropriate limit of a single perfect conductor.

The first term on the right-hand side in the above expression is just the usual photon propagator. The correction comes from the second one but since it does not depend only on the distance between two points, but also on the location of the mirror, this anisotropy prevents us from transforming the propagator to momenta space as usual. In order to overcome this difficulty and verify that this new propagator enjoys the desired properties, we can resort to classical external sources to describe pointlike charges \cite{BaroneHidalgo1} instead of scattering methods.

\section{The Image Method}
\label{image}

In this section we show how the above propagator leads to the correct interaction between a static charge and the plane mirror, exploiting the expression for the total energy of the system in terms of the functional generator of connected Green functions, i.e.
\begin{equation}
\label{EQ12}
E=\lim_{T\rightarrow\infty}\frac{1}{2T}\int d^{4}x\int d^{4}y\ J^{\mu}(x)G_{\mu\nu}(x,y)J^{\nu}(y)\ .
\end{equation}

The presence of a pointlike charge is accomplished by the external source,
\begin{equation}
\label{EQ13}
J^{\mu}(x)=q  \eta^{\mu 0} \delta^{(3)}(\textbf{x}-\textbf{b})\ .
\end{equation}
where $\textbf{b}$ is a constant vector standing for the charge position that will be taken to be $\textbf{b}=(0,0,b)$ for the sake of simplicity.

Substituting the current (\ref{EQ13}) in (\ref{EQ12}) and integrating the delta functions yields,
\begin{equation}
\label{EQ14}
E=\lim_{T\rightarrow\infty}\frac{q^{2}}{2T}\int^{T/2}_{-T/2} dx^{0} \int \frac{d^{2} \textbf{p}_\parallel}{(2\pi)^{2}}\  \mathcal G_{00}(b,b;p^{0}=0,\textbf{p}_\parallel) \ .
\end{equation}

Noting that the first term on the right-hand side of Eq.(\ref{EQ11}) is just the free propagator and therefore does not contribute to the interaction energy between the mirror and the charge \footnote{This term does not depend on the distance between the mirror and the charge. It is present even in the absence of the mirror and gives the charge self energy.}, we can use the remaining part of the propagator to write Eq.(\ref{EQ14}) as,
\begin{equation}
\label{EQ15}
E_{int}=-\frac{q^{2}}{4} \int \frac{d^{2} \textbf{p}_\parallel}{(2\pi)^{2}}\ \frac{e^{-2\sigma |b-a|}}{m+\sigma} \ .
\end{equation}
The above expression is the form of the interaction energy between a static charge $q$, located at $\textbf{b}=(0,0,b)$, and the mirror. Defining $R=|b-a|$, changing to polar coordinates and using the differential operator, we obtain
\begin{eqnarray}
\label{EQ16}
E_{int}(R,m)&=&-\frac{q^{2}}{4} \int \frac{dr d\theta}{(2\pi)^{2}}\ \frac{re^{-2rR}}{m+r} \cr\cr
&=& \frac{q^{2}}{16\pi} \frac{\partial}{\partial R} \ \int^{\infty}_{0} dr \ \frac{e^{-2rR}}{m+r} \ .
\end{eqnarray}

This expression can be brought to a convenient form by performing the change of variable $s=r+m$, so that
\begin{eqnarray}
\label{EQ17}
E_{int}&=&\frac{q^{2}}{16\pi} \frac{\partial}{\partial R} \Biggl(\ e^{2mR} \ \int^{\infty}_{m} ds \ \frac{e^{-2sR}}{s} \Biggr) \ .
\end{eqnarray}

The integral above is the well-known exponential integral function $Ei(u,v)$ \cite{Arfken}. So, we can write the energy as,
\begin{eqnarray}
\label{EQ18}
E_{int}(R,m)&=&\frac{q^{2}}{16\pi} \frac{\partial}{\partial R} \ [ e^{2mR} \ Ei(1,2mR) ] \cr\cr
&=&-\frac{q^{2}}{16\pi R}[1-2mR e^{2mR} \ Ei(1,2mR)] \ .\cr
&\ &\ 
\end{eqnarray}
Eq.(\ref{EQ18}) is the exact result for the interaction energy of a point charge and a partially reflective surface, described by the model (\ref{EQ1}). The second term in the second line accounts for the correction due to the partial reflectivity. This term vanishes in the limit $m\to0$, which corresponds to the field subjected to boundary conditions imposed by a perfectly conducting plate. In this case the energy (\ref{EQ18}) reads
\begin{eqnarray}
\label{EQ19}
\lim_{m\to 0}E_{int}(R,m)=-\frac{q^{2}}{16\pi R} \ .
\end{eqnarray}
Therefore, the lagrangian (\ref{EQ1}) describes exactly the interaction obtained via image method in the limit of a perfect mirror. In the opposite case, when $m\to\infty$, the energy (\ref{EQ18}) goes to zero as expected. Also, for a fixed distance $R$, it decreases as $m$ increases. 

\begin{figure}[!h]
 \centering
   \includegraphics[scale=0.35]{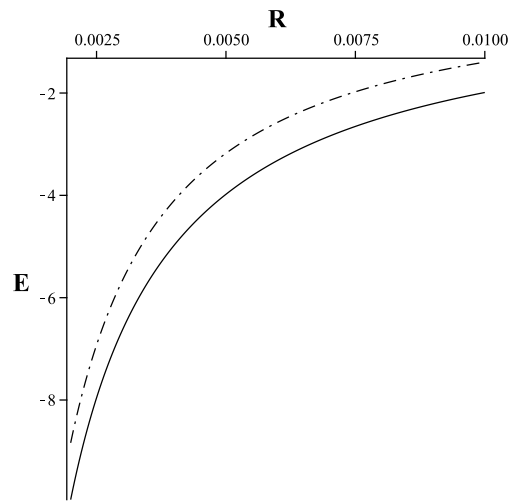}
   \caption{Interaction energy as a function of the distance, Eq.(\ref{EQ18}), for $q=1$, $m=10$ (dashed line) and $m=0$ (solid line).}
  \label{fig1}
\end{figure}

The behavior of the energy as a function of the distance $R$ can be seen in Fig.(\ref{fig1}), where we show a plot of Eq.(\ref{EQ18}) for two different values of $m$.

The force between the mirror and the charge is;
\begin{eqnarray}
F&=&-\frac{\partial}{\partial R}E_{int}(R,m)\cr\cr
&=&-\frac{q^{2}}{16\pi R}\Bigl[1-2mR+(2mR)^{2}e^{2mR}Ei(1,2mR)\Bigr] .\cr
&\ &\ 
\end{eqnarray}
Note that this expression is monotonic in $R$ and $m$, and is always negative, which denotes an attractive force.

\section{Final remarks}
\label{final}

In conclusion, we proposed an enlarged Maxwell lagrangian amended by a new gauge invariant term, and featured by a free parameter to gauge the degree of transparency of a $\delta$-like surface, with the optical properties of an uniaxial dielectric. We also found the modification undergone by the photon propagator due to it, and computed the interaction energy between a static charge and the semitransparent mirror. It turns out that, in the appropriate limit, the obtained propagator reduces to the one for a perfect conducting plate. This allowed us to show that the interaction between a static charge and a perfectly reflective surface that arises from this lagrangian is exactly the same as the one obtained via the classical image method in the limit of a perfect mirror. 

This effective lagrangian is necessary for a field theoretic treatment of the electromagnetic field in the presence of a partially reflective surface, since the usual methods based on external potentials coupled to fields, widely used to study the behavior of the scalar and fermionic fields under boundary conditions, cannot be employed without destroying the gauge invariance. Furthermore, if the lagrangian (\ref{EQ1}) can lead directly to the Casimir effect in setups where realistic properties must be taken into account, is another point of interest that must be addressed, although it is not a simple one \cite{andamento}.

Moreover, since $\delta$-like external sources can describe charges and mulipole distributions along branes of arbitrary dimensions in several different scenarios \cite{BaroneHidalgo1,BaroneHidalgo2,BBH}, we hope that the model presented here can advance the endeavor of better understanding some interesting physical signatures. Not only about the long-standing issue of atoms in interaction with different kinds of surfaces but also regarding the interaction between dissimilar surfaces and multilayered materials (see for instance \cite{graphene}).

\begin{acknowledgments}
The authors would like to thank CNPq and CAPES (brazilian agencies), for financial support.
\end{acknowledgments}



\end{document}